\documentclass[aps,prb,showpacs,floatfix,twocolumn,superscriptaddress]{revtex4-1}
\usepackage{graphicx}    
\usepackage{hyperref}    
\usepackage{bm}          
\usepackage[sumlimits,intlimits]{amsmath}
\usepackage{amsfonts,amssymb,mathtools}
\usepackage{tikz}
\usepackage{cancel}
\usetikzlibrary{arrows, decorations.pathmorphing,patterns,snakes}
\pdfoutput=1

\begin{document}

\title{Excitonic Collapse in Semiconducting Transition Metal Dichalcogenides}

\author{A. S. Rodin}
\affiliation{Boston University, 590 Commonwealth Ave., Boston MA 02215}

\author{A. H. Castro Neto}
\affiliation{Boston University, 590 Commonwealth Ave., Boston MA 02215}
\affiliation{Graphene Research Centre and Department of Physics, National University of Singapore, 117542, Singapore}

\date{\today}
\begin{abstract}
Semiconducting transition metal dichalcogenides (STMDC) are two-dimensional (2D) crystals characterized by electron volt size band gaps, spin-orbit coupling (SOC), and d-orbital character of its valence and conduction bands. We show that these materials carry unique exciton quasiparticles (electron-hole bound states) with energy within the gap but which can ``collapse'' in the strong coupling regime by merging into the band structure continuum. The exciton collapse seems to be a generic effect in these 2D crystals.    
\end{abstract}

\maketitle

\section{Introduction}
\label{sec:Introduction}

Isolation of graphene in 2004 \cite{Novoselov2004efe} opened up a new field in condensed matter physics: 2D crystals 
\cite{Novoselov2005tda}. Having found a stable 2D crystal, the scientific community began searching for 
other 2D systems with different properties \cite{CastroNeto2011ndi}. A particularly interesting family of materials is the 
transition metal dichalcogenides (TMDC) with chemical formula {\it M}{\it C}$_2$, where {\it M} is a transition
metal and {\it C} a chalcogen ({\it C} = S, Se,Te) \cite{Wilson1975cdw,Wang2012eao,Chhowalla2012tco}. At their thinnest, these systems are composed of 
three atomic layers: one layer of {\it M} atoms is sandwiched by two {\it C} atom layers. In each layer, atoms 
form a triangular lattice so that the material resembles graphene where instead of A and B sublattices 
one has different types of atoms found in different planes. These materials show a large class
of ground states that vary from metallic or semiconducting to superconducting or charge density wave (CDW) 
depending on the {\it M} atom.

Within this family, there are several members that are semiconducting (MoS$_2$, MoSe$_2$, WS$_2$, WSe$_2$, etc.)
with a band gap with energy in the visible frequency range, between 1 eV to 3 eV, and with striking
electronic \cite{Fuhrer2013mom} and optical \cite{Mak2010atm} properties. Hence, because of their
ultimate thinness and softness these materials can play an important technological role in flexible electronics and
flexible photovoltaics, among others. Moreover, due to the d-nature of the orbitals, the valence and conduction
bands are characterized by a large density of states. Aditionally, with the presence of 
heavy {\it M} atoms, the SOC can be enhanced allowing for unique physical phenomena in a truly 2D crystal.

A fascinating aspect of the band structure of several semiconducting TMDC's (STMDC) is that the low-energy behavior, close to the
band edges, can be described by a ``massive'' Dirac Hamiltonian where the band gap is proportional to the mass 
of the electrons \cite{Xiao2012csa}. In this paper, we study the dielectric screening of these unique systems and 
the binding of electrons and holes into  exciton quasiparticles (EQ). The energy of EQ resides within the gap 
of the STMDC but in the strong coupling regime (to be defined below) we find that EQs can actually merge 
with the valence (conduction) continuum leading to something akin to the ``atomic collapse'' or ``fall to the center'' 
effect. This phenomenon, predicted for heavily charged impurities in graphene in 2007 \cite{Pereira2007cip,Shytov2007aca}, 
was only observed recently \cite{Wang2013oac}. We show that the same situation can occur in STMDC without the need 
of impurities but as a result of the internal electron-hole interaction in these materials. Note that the numerical values used in this text are taken from Ref.~\onlinecite{Xiao2012csa}. While these quantities vary in literature, our primary task is to demonstrate the behavior for the physically attainable gap and spin-orbit coupling.

This paper is structured as follows. The model is described in Sec.~\ref{sec:Model}. Next, we determine the screening in the system by computing the polarization function in Sec.~\ref{sec:Polar}. The main calculations of the text concerning the excitonic collapse is contained in Sec.~\ref{sec:Collapse}. Finally, we summarize our finding in Sec.~\ref{sec:Conclusion}.

\section{Model}
\label{sec:Model}

The starting point of our discussion is the low-energy Hamiltonian used to model the band structure close to 
$\mathrm{K}$ and $\mathrm{K}'$ points in the Brillouin zone. Using arguments based on symmetry, the authors of Ref.~\onlinecite{Xiao2012csa} proposed the following form
\begin{equation}
H = at(\tau k_x\sigma_x+k_y\sigma_y)+\frac{\Delta}{2}\sigma_z-\lambda\tau\frac{\sigma_z-1}{2}s_z\,,
\label{eqn:H}
\end{equation}
where $\tau$ is the valley index, $a$ is the lattice constant, $t$ is the hopping energy, $\Delta$ is the ``mass'' 
term,  $\sigma_i$ is the Pauli spin matrix, $s_z$ is a diagonal $4\times 4$ matrix with the diagonal $(1, 1,-1,-1)$, and $\lambda$ is the SOC parameter. This simple, yet powerful Hamiltonian was introduced in Refs.~\onlinecite{Kane2005qsh,Kane2005zto} in the context of graphene. The first terms is the familiar linear dispersion at the corners of the Brillouin zone. The second term of Eq.~\eqref{eqn:H} is used to describe staggered lattice potential in graphene; the third term captures the spin-orbit interaction. Since the underlying lattice symmetry is the same for TMDC's, this Hamiltonian is used to describe these materials. It is important to keep in mind, however, that unlike graphene, where the spinor components describe $p_z$ orbitals of $A$ and $B$ sublattices, here the spinor labels the conduction and valence bands. The conduction band is primarily composed of $d_{z^2}$ orbitals and the valence is based on $d_{x^2-y^2}$ and $d_{xy}$.~\cite{Xiao2012csa} The difference of orbital energies results in the mass term.

 In this model, the masses of electrons and holes are set to be equal and trigonal warping is neglected \cite{Rostami2013elh}. These effects can be added to Eq.~\eqref{eqn:H}, but they do not affect qualitatively the results here. The values for the constants in Eq.~\eqref{eqn:H} are given in 
Ref.~\onlinecite{Xiao2012csa}. The Hamiltonian in Eq.~\eqref{eqn:H} is block-diagonal and the spins are uncoupled. 
This allows us to treat the spin sub-Hamiltonians separately, leading to a simplified $2\times2$ matrix:
\begin{equation}
H_{\uparrow/\downarrow} =\tau\left[
 \begin{pmatrix}
m_l&take^{-i\tau\theta}
\\
take^{i\tau\theta}&-m_l
\end{pmatrix}
\pm\frac{\lambda}{2}\mathbb{I}\right] -\mu\mathbb{I}= \tau H_{0,l}-\mu^{*}_l \mathbb{I}
\label{eqn:H_red}
\end{equation}
where $\mu$ is the chemical potential (which can be tuned by an applied electric field), 
$m_\pm = (\Delta\mp\lambda\tau)/(2\tau)$, and $\mu^{*}_\pm = \mu\mp\tau\lambda/2$. 
Equation~\eqref{eqn:H_red} is the Hamiltonian of a ``massive'' Dirac fermion with a finite chemical potential 
where both $\mu^{*}_l$ and $m_l$ depend on valley and spin indices. 

The mass term in Eq.~\eqref{eqn:H} breaks the ``sublattice" symmetry leading to a gap of size $2 m_l$ between valence
and conduction bands. The conduction, $s=+1$, and valence, $s=-1$, bands are given by 
$\varepsilon_{k,l,s} = s \sqrt{m_l^2+(tak)^2}$, revealing the emergent low energy Lorentz invariance of these 
materials close to the band edge. Furthermore, the spin-up and spin-down Green's functions are given by:
\begin{align}
\mathcal{G}_{\uparrow/\downarrow} &= \left[i\hbar\omega_n-H_{\uparrow/\downarrow}\right]^{-1} 
= \frac{1}{2}\sum_{s = \pm1}\frac{\mathbb{I}+s\tau H_{0,l}/\varepsilon_{k,l}}{i\hbar\omega_n-\xi_{k,l,s}}\,,
\label{eqn:G}
\\
\xi_{k,l,s} &= s \varepsilon_{k,l}-\mu^{*}_l \,,\quad \varepsilon_{k,l} = \sqrt{m_l^2+(tak)^2}\,.
\label{eqn:xi_eps}
\end{align}

\section{Polarization Function}
\label{sec:Polar}

\begin{figure}[h]
\includegraphics[width = 3.0in]{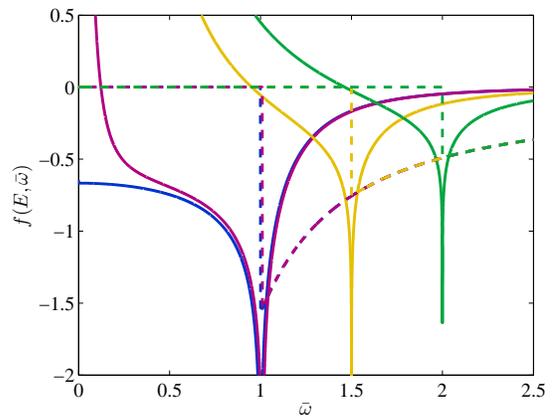}
\caption{(Color online) Real (solid) and imaginary (dashed) plots of $f(E,\bar\omega) = 8\pi|m_l| P(q,\omega)/q^2$ 
from Eq.~\eqref{eqn:P_q} for $E_l = 1$ (blue), $1.01$ (purple), 1.5 (yellow), and 2 (green).}
\label{fig:Static_Pol}
\end{figure}

In order to understand the electron-electron interactions, which eventually will lead to the exciton formation, it is important to study the behavior of the dielectric function. In the random phase approximation (RPA), the polarization function 
is given by the density-density correlation:
\begin{equation}
P_l(q,i\omega_n) = \mathrm{Tr}\left[\sum_{k,ik_n}\mathcal{G}(k,ik_n)\mathcal{G}(k+q,ik_n+i\omega_n)\right]\,.
\label{eqn:P}
\end{equation}
In the low temperature regime, $T \to 0$, and long wavelength limit, $q \to 0$, we find: 
\begin{equation}
P_l(q\rightarrow0,\omega)=\frac{q^2}{8\pi|m_l|}\left[\frac{E_l}{2\bar\omega_l^2}-\frac{1+\bar\omega_l^2}{4\bar\omega_l^3}
\ln\left(\frac{E_l+\bar\omega_l}{E_l-\bar\omega_l}\right)\right]\,,
\label{eqn:P_q}
\end{equation}
where $E_l = \text{max}\left[\mu^*_l/m_l,1\right]$, $\bar\omega_l = \hbar\omega/(2|m_l|)$, and $\mu^{*}_l>0$. 
For a vanishing gap, $2|m_l| \to 0$, the result approaches the graphene polarization function \cite{Kotov2012eei}. 
One important difference is the fact that the imaginary part of the product $\bar\omega_l P(q,\bar\omega_l)$, 
proportional to the low-$q$ conductivity, is not constant for $\bar\omega_l>E_l$. Instead, it has a maximum at 
$E_l = \bar\omega_l$ which is the consequence of the high density of states at the extremal points of the bands. 
For undoped or ungated crystals, the chemical potential is located in the gap region, the conduction band is empty, 
and the conductivity becomes finite only for $\hbar\omega_l = 2|m_l|$. As the system becomes n-doped, $\bar\omega_lP(q,\bar\omega_l)$ is nonzero at $\hbar\omega_l > 2\mu^{*}_l$ since interband transitions with smaller $\bar\omega_l$ are forbidden. We plot the polarization function for several values of $E_l$ in Fig.~\ref{fig:Static_Pol}. If the Fermi level is in the gap, the real part of polarization is always negative. This means that the dielectric function $\epsilon_\mathrm{RPA}(q,\omega)=\epsilon_0-2\pi e^2/q\times P(q,\omega)$, where $\epsilon_0$ is the external dielectric constant, never changes sign and there are no collective charge 
excitations such as plasmons. Conversely, allowing electrons to populate the conduction band allows 
$\epsilon_\mathrm{RPA}$ to pass through zero, giving rise to plasmonic behavior.

The screened Coulomb interaction between charge particles is given by the static polarization function. 
Since we are interested in excitons, we treat the case where the Fermi level 
is in the middle of the gap. Straightforward manipulations of Eq.~\eqref{eqn:P} at $T = 0$ yield:

\begin{widetext}
\begin{align}
P(q,0) & =-\frac{m_l}{4\pi^2(ta)^2}F\left(\frac{taq}{m_l}\right)\,,
\label{eqn:P_omega}
\\
F(Q)&=\oint d\theta\int_0^\infty dx\,x\left(1-\frac{1+x(x+Q\cos\theta)}{\sqrt{1+x^2}\sqrt{1+x^2+Q^2+2 xQ\cos\theta}}\right)\left(\frac{1}{\sqrt{1+x^2}+\sqrt{1+x^2+Q^2+2 xQ\cos\theta}}\right)\,.
\label{eqn:F_Q}
\end{align}
\end{widetext}
The asymptotic behavior of $F(Q)$ is given by:
\begin{equation}
F(Q) = \begin{dcases}
\frac{\pi Q^2}{3}\,\mathrm{if}\,Q\ll1\,,\\
\frac{\pi^2 Q}{4}\,\mathrm{if}\,Q\gg1\,.
\end{dcases}
\end{equation}
The full system polarization is obtained by doubling Eq.~\eqref{eqn:P_omega} to account for the valley degeneracy and summing over $l = \pm1$ to take 
care of both spins. Dividing the Coulomb term $2\pi e^2/q$ by the static $\epsilon_\mathrm{RPA}$ gives the screened 
interaction potential. Taking the inverse Fourier transform of $V_\mathrm{RPA}(q)$ results in:
\begin{align}
V_\mathrm{RPA}(\rho a) &= \frac{\hbar v \alpha}{\rho a}\int_0^\infty dx\frac{J_0\left(x\right)}{1+\frac{\alpha}{\pi}\sum_{l=\pm1}\frac{F[x/(S_l^2\rho )]}{x/(S_l^2\rho)}}\,,
\label{eqn:V_RPA}
\\
\alpha &= \frac{e^2}{\hbar v\epsilon_0}\,,\quad S_l = \sqrt{\frac{m_l}{t}}\,,
\end{align}
where $\hbar v = ta$. For very small $r$, we can use the large-$Q$ approximation for $F(Q)$ in Eq.~\eqref{eqn:V_RPA}. 
This removes all $x$-dependence from the denominator in \eqref{eqn:V_RPA} and results in a simple renormalization of 
the coupling constant $\alpha\rightarrow\alpha/(1+\alpha\pi/2)$ which is identical to the static screening in undoped single 
layer graphene \cite{Kotov2012eei}. In the opposite limit of large $r$, there is no such renormalization and the Coulomb 
interaction remains unscreened.

To understand the nature of the screening, it is convenient to rewrie the polarization as
\begin{equation}
P(q,0) = -\frac{m_l}{4\pi^2(ta)^2}\left[\left(F\left(\frac{taq}{m_l}\right)-\frac{\pi^2}{4}\frac{taq}{m_l}\right)+\frac{\pi^2}{4}\frac{taq}{m_l}\right]\,.
\label{eqn:Pol_Split}
\end{equation}
Next, we can obtain the charge induced by an external potential $\Phi_\text{ext}(r) = e/(\epsilon_0 r)$, with the Fourier transform $\Phi_\text{ext}(q) = 2\pi e/(\epsilon_0 q)$. Using the simplest linear responce, one has $n(q) = e\Phi_\text{ext}(q)P(q)$. The second term in Eq.~\eqref{eqn:Pol_Split} yields
\begin{equation}
n(r) = -\frac{\pi\alpha}{8}\int dq\,qJ_0(qr) = -\frac{\pi\alpha}{8}\frac{\delta(r)}{r}\,.
\end{equation}
Integrating over $r$ in polar coordinates and multiplying by four to account for the degeneracy yields $-\pi\alpha/2$, located at the origin. This means that the external charge $e$ induces a screening charge $-e \pi\alpha/2$ located virtually at the same position. In fact, rewriting the screened energy at small $r$ as
\begin{equation}
V_\text{RPA}(r) = \frac{\hbar v \alpha}{r}\sum_{n = 0}\left(-\frac{\pi\alpha}{2}\right)^n
\end{equation}
shows that the screening arises from a number of induced charges with alternating signs located at the origin.

The first term in Eq.~\eqref{eqn:Pol_Split} has the sign opposite to the second term. It means that the charges induced from it are also opposite. Since the second term results in localized screening charges, those arising from the first term are compensating anti-screening. They are not localized and their effect becomes more pronounced at larger $r$ as the total sum enclosed in $r$ grows, eventually matching the screening one. A similar result was previously obtained for gapped graphene.~\cite{Kotov2008pcd} The authors showed that the anti-screening charge density has a logarithmic decay at smaller $r$ and drops off as $r^{-3}$ at large distances. It was also demonstrated that most of the anti-screening charge is located at $r<2\hbar v/\Delta = 2ta/\Delta$. Given that, according to Ref.~\onlinecite{Xiao2012csa}, $t/\Delta<1$, the interaction becomes unscreened within a few lattice constants. Because of this, we will use the unscreened version, keeping in mind that the coupling constant $\alpha$ may need to be adjusted from $e^2/\hbar v\epsilon_0$ to include the screening at small enough distances.

\section{Excitonic Collapse}
\label{sec:Collapse}

Having established the nature of the Coulomb interaction in the system, we now turn our attention to the EQ. 
In order to keep the notation as simple as possible, we will dispense with the details of the SOC and simply 
denote the size of the gap by $2m_l$, keeping in mind that this quantity depends on the spin and the valley. 
It is convenient to perform a particle-hole transformation so that the electron and hole free Hamiltonians become:
\begin{equation}
H_{e/h} = \begin{pmatrix}
 \pm m_l&\pm\hbar v k e^{\mp i\theta}
\\
\pm\hbar v k e^{\pm i \theta}&\mp m_l
\end{pmatrix}\,.
\label{eqn:H_eh}
\end{equation}
Diagonalizing the Hamiltonian gives two branches with $\mathcal{E} = \pm\sqrt{m_l^2+\hbar^2v^2k^2}$. Positive energies 
correspond to the states in the electron and hole bands while the negative ones represent the states in the electron and 
hole ``seas". The total Hamiltonian for two particles without interactions is given by the 
tensor product $H_2 = H_e\otimes\mathbb{I}+\mathbb{I}\otimes H_h$ \cite{Sabio2010tbp}:
\begin{equation}
H_2 = \begin{pmatrix}
0&-\hbar vke^{i\theta_k}&\hbar vpe^{-i\theta_p}&0
\\
-\hbar vk e^{-i\theta_k}&2m_l&0&\hbar vpe^{-i\theta_p}
\\
\hbar vpe^{i\theta_p}&0&-2m_l&-\hbar vke^{i\theta_k}
\\
0&\hbar vpe^{i\theta_p}&-\hbar vke^{-i\theta_k}&0
\end{pmatrix}\,.
\label{eqn:H2}
\end{equation}
When dealing with excitons, one works in the center-of-mass frame of reference. Since in the model the masses of 
electrons and holes are the same, the momenta must be opposite. Setting $p = -k$ results in four eigenvalues: 
$\pm2\sqrt{m_l^2+\hbar^2v^2k^2}$ and a doubly degenerate $0$. The zero energy eigenstates arise from the cases 
when the system has a single electron or a hole and its complementary particle is in its sea. This way, since 
they have the same momentum, they give equal and opposite contributions to the total energy. The negative 
eigenvalue corresponds to the situation where both the electron and the hole are in their respective seas. 
Finally, the positive eigenvalue is what we are interested in. There, an electron is found in the electron 
band and the hole is the hole band. This can be regarded as an excitonic state with a vanishingly weak interaction. 
Therefore, the states of interest constitute a subspace of a full Hilbert space describing a two-particle system. 

If we go back to the laboratory frame and investigate the kinetic energy of the exciton, $k\rightarrow K/2-q$ and $p\rightarrow K/2+q$. Here, $q$ is the motion in the center of mass frame and $K$ is the momentum of the center of mass. Diagonalizing Eq.~\eqref{eqn:H2}, one cannot separate the two momenta. However, assuming that $q\ll K$, it is possible to estimate the kinetic energy as $\mathcal{E}_K\approx2\sqrt{m_l^2+\hbar^2v^2K^2/4}$.

In principle, one can keep the masses of electrons and holes different. A particular case of interest is when 
one of these masses goes to infinity. In this case, the problem reduces to a particle moving in the field of 
a fixed charged impurity. Careful expansion of the two-particle eigenvalues gives $\sqrt{m_l^2+\hbar^2v^2k^2}$ for 
the excitonic energy, where $m_l$ is finite. The collapsing states of this system were treated in 
Ref.~\onlinecite{Pereira2008sci}.

\begin{figure}[t]
\includegraphics[width = 3in]{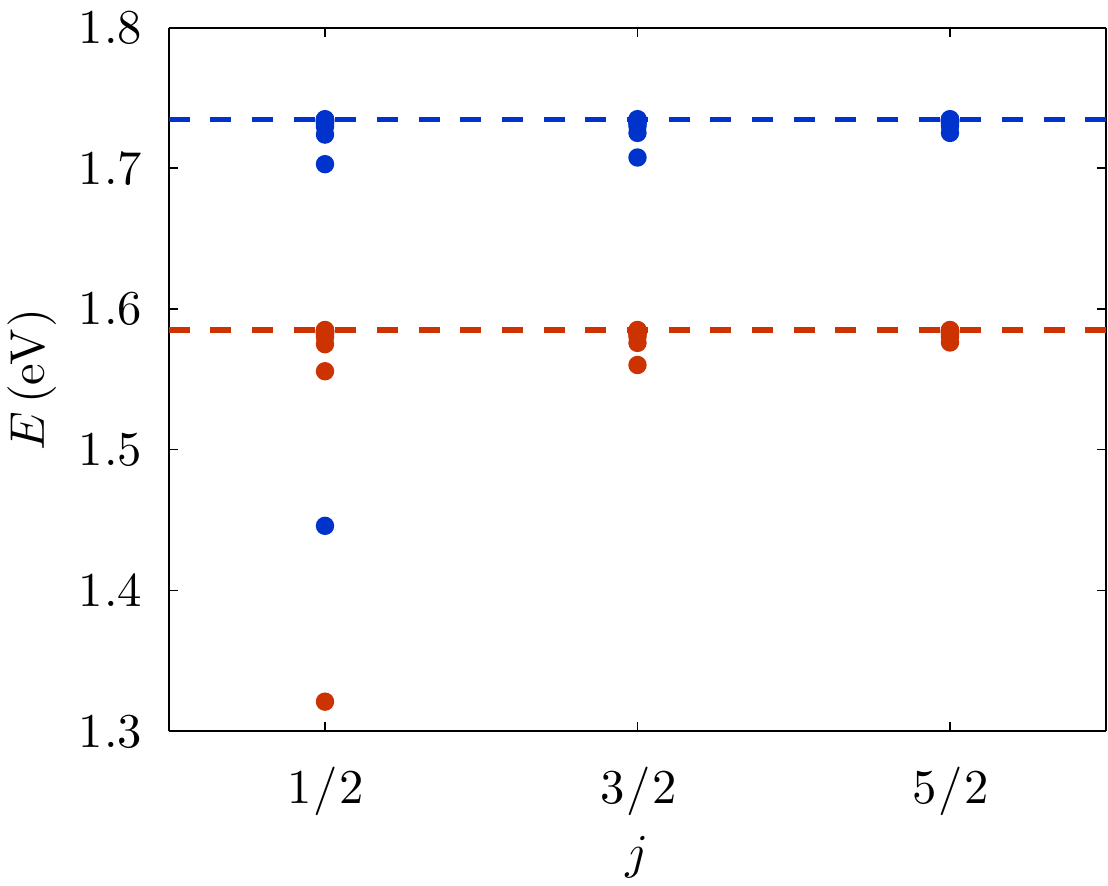}
\caption{(Color online) Optical absorption energies for $\mathrm{MoS}_2$ ($a = 3.193\,\text{\AA}$, $\Delta = 1.66$ eV, $t = 1.1$ eV, $\lambda = 0.075$ eV) on SiC. Red and blue circles correspond to different spins due to SOC. The dashed lines denote the energies equal to the band gaps for both spins. As the principal quantum number $n\rightarrow\infty$, the absorption energy approaches $\Delta\pm\lambda$, depending on the spin.}
\label{fig:MoSe2_Levels}
\end{figure}

In the case of equal masses, the minimum energy of exciting an electron to the conduction band from the valence 
band is $2m_l$. We set up the bands in such a way that zero energy is located half-way between their extrema. 
This results in a reduced excitonic Hamiltonian:
\begin{equation}
H_E = \begin{pmatrix}
m_l&2qe^{-i\theta}
\\
2qe^{i\theta}&-m_l
\end{pmatrix}-\frac{\hbar v \alpha}{r}\,,
\label{eqn:HE}
\end{equation}
where we have included the interaction term. Note that compared to the impurity problem~\cite{Pereira2008sci}, the momentum is doubled
since the electron and hole are moving with the same momentum. We can perform a variable transformation 
$2q\rightarrow p$, $r\rightarrow2x$ to preserve the commutation relation, yielding:
\begin{equation}
\bar H_E = \begin{pmatrix}
m_l&pe^{-i\theta}
\\
pe^{i\theta}&-m_l
\end{pmatrix}-\frac{\hbar v \bar\alpha}{x}
\label{eqn:HEbar}
\end{equation}
with $\bar\alpha = \alpha/2$. Notice that for the exciton problem the effective fine structure constant $\bar\alpha$ 
is half of the impurity case because of the doubling in the momentum. Equation (\ref{eqn:HEbar}) describes a massive 
Dirac particle in an attractive Coulomb potential of a single charge with coupling that is $2c/(v\kappa)$ times 
stronger than the vacuum fine structure constant. The energies of the bound states are given by \cite{Pereira2008sci}:
\begin{equation}
\mathcal{E}_{jn} =m_l\frac{n+\sqrt{j^2-\bar\alpha^2}}{\sqrt{\bar\alpha^2+\left[n+\sqrt{j^2-\bar\alpha^2}\right]^2}}\,,
\label{eqn:Energies_bound}
\end{equation}
where $n$ is the principal quantum number and $j = \pm1/2,\,\pm3/2\dots$ is the angular momentum quantum number.
These determine the binding energies for the EQ and can be measured, say, by optical means. The optical absorption 
energies are given by the sum of $m_l$ and Eq.~\eqref{eqn:Energies_bound} since as $n\rightarrow\infty$, this sum 
approaches the size of the gap. Notice that for $\bar\alpha>1/2$, some energies become imaginary. This is caused by 
the breakdown of the point-charge treatment of the center of the Coulomb well and is usually remedied by a 
regularization procedure, as we discuss below. For now, we avoid this problem by choosing $\bar \alpha<1/2$ to illustrate a particular case. As an example, we pick $\mathrm{MoS}_2$ on a SiC substrate ($\epsilon_0\approx 5.5$ and $\bar \alpha\approx 0.42$)\cite{Xiao2012csa}. Fig.~\ref{fig:MoSe2_Levels} shows the absorption energies for this system. It is important to keep in mind that the exact system parameters depend strongly on experimental conditions. The nature of the substrate and the quality of the sample play a role. This means that these parameters need to be determined on a case-by-case basis.

For a regular Coulomb potential, the energies in Eq.~\eqref{eqn:Energies_bound} for the states are always positive. At sufficiently large $\bar\alpha$, however, the energy ``dives'' to negative value until it eventually merges with the valence 
band at $\mathcal{E} =-m$. To analyze this phenomenon, we need to introduce a regularization procedure to 
deal with the singularity of the Coulomb interaction. This entails imposing a potential cutoff at $x = a/2$ 
above which the interaction retains its usual $1/x$ form and for $x<a/2$ it becomes $-2\hbar v\bar\alpha/a$. The reason for choosing this particular length is that the physical lengths in the system are limited by the lattice spacing.

The form of the solution to Eq.~\eqref{eqn:HEbar} that we impose is:
\begin{equation}
\Psi_j(x,\theta) = \frac{1}{\sqrt{x}}\begin{pmatrix}
e^{-i(j-1/2)\theta}A(x)\\
ie^{-i(j+1/2)\theta}B(x)
\end{pmatrix}\,.
\label{eqn:Spinor}
\end{equation}
Setting $\mathcal{E}\rightarrow- m$ and solving Eq.~\eqref{eqn:HEbar} for $x>a/2$ yields \cite{Pereira2008sci}:
\begin{align}
A^>(x) &= \mathcal{N}K_{i\nu}\left(\sqrt{\frac{8\bar\alpha m_l x}{\hbar v}}\right)\,,
\label{eqn:A>}
\\
B^>(x) &= \frac{\mathcal{N}}{\bar\alpha}\left[\left(j+\frac{i\nu}{2}\right)K_{i\nu}\left(\sqrt{\frac{8\bar\alpha m_l  x}{\hbar v}}\right)+\right.
\nonumber
\\
&\left.\sqrt{\frac{2\bar\alpha m_l x}{\hbar v}}K_{i\nu-1}\left(\sqrt{\frac{8\bar\alpha m_lx}{\hbar v}}\right)\right]\,,
\label{eqn:B>}
\end{align}
where $\nu = 2\sqrt{\bar\alpha^2-j^2}$, $K_n$ is a modified Bessel function, and $\mathcal{N}$ is the normalization constant. Similarly, for $x<a/2$, the result becomes \cite{Pereira2008sci}:
\begin{align}
A^<(x) &= \mathcal{N}'\sqrt{x}\bar\alpha J_{j-1/2}(kx)\,,
\label{eqn:A<}
\\
B^<(x) &= \mathcal{N}'\sqrt{x}\frac{ka}{2} J_{j+1/2}(kx)
\label{eqn:B<}
\end{align}
with $k = 2\sqrt{\bar\alpha^2-S_l^2\bar\alpha}/a$. The solutions need to be matched at the boundary so that:
\begin{equation}
\left. A^<(\bar\alpha)/B^<(\bar\alpha)\right|_{x = a/2}=\left. A^>(\bar\alpha)/B^>(\bar\alpha)\right|_{x=a/2}\,,
\label{eqn:matching}
\end{equation}
resulting in:
\begin{align}
&\frac{J_{j-1/2}(ka/2)}{\sqrt{\bar\alpha^2-S_l^2\bar\alpha} J_{j+1/2}(ka/2)}=
\nonumber
\\
&\left[\left(j+\frac{i\nu}{2}\right)
+S_l\sqrt{\bar\alpha}K_{i\nu-1}\left(2S_l\sqrt{\bar\alpha}\right)/K_{i\nu}\left(2S_l\sqrt{\bar\alpha}\right)\right]^{-1}\,.
\label{eqn:alpha_sol}
\end{align}

The cutoff point $a/2$, while based on a physical quantity, might appear somewhat arbitrary. If one were to allow the cutoff to approach infinity, the critical coupling $\bar\alpha$ would approach the lowest root of $J_{j-1/2}(\sqrt{\bar\alpha^2-S_l^2\bar\alpha})$. Therefore, the true value lies between these two limiting cases. In reality, one expects the cutoff to be closer to $a/2$ than to infinity.

\begin{figure}[t]
\includegraphics[width =3in]{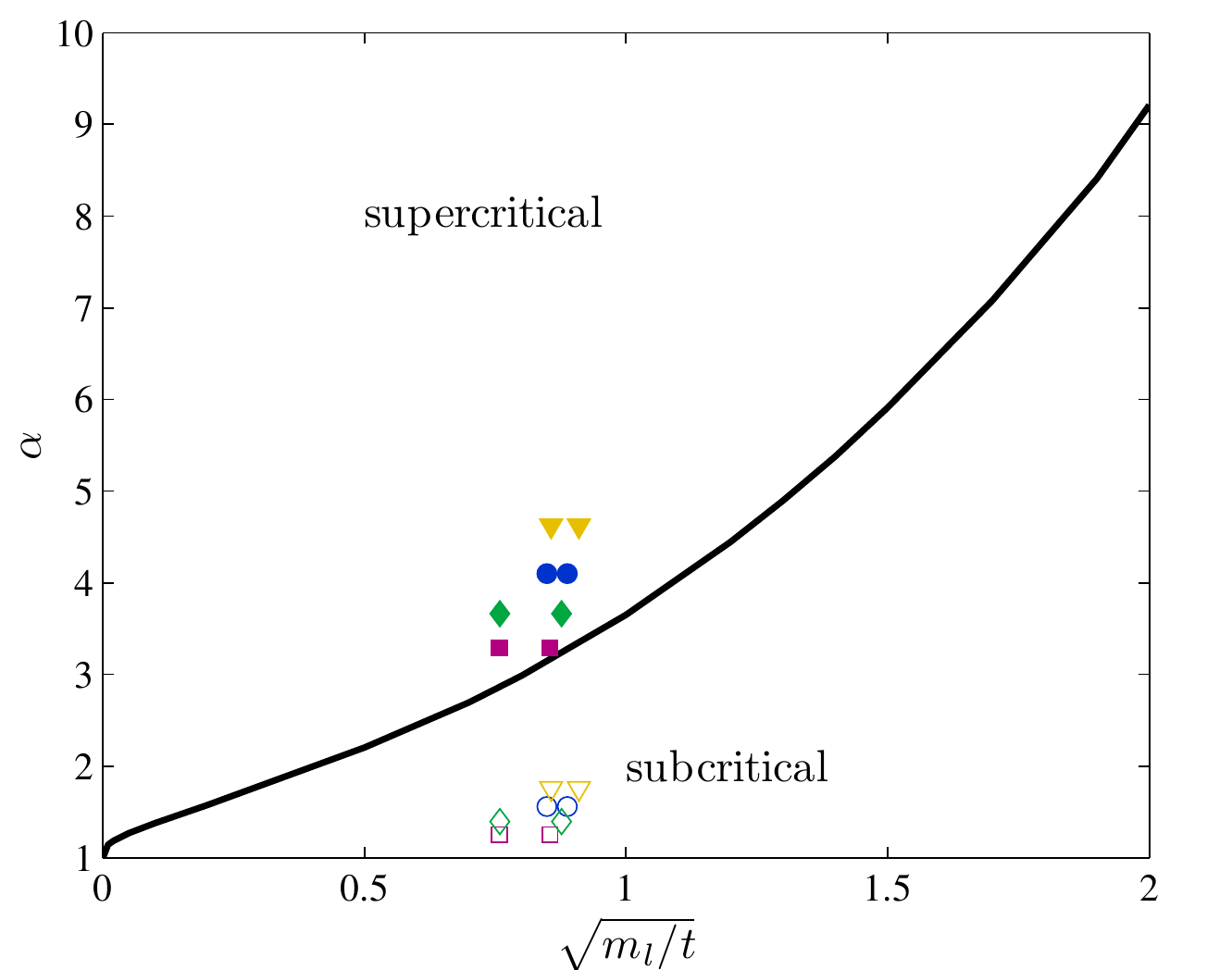}
\caption{(Color online) A phase diagram obtained from the solution of Eq.~\eqref{eqn:alpha_sol} for a range of $S_l = \sqrt{m_l/t}$. 
We also show where STMDS are located on this plot if they are on vacuum, $\epsilon_0 = 1$ (filled symbols), and if the samples are 
placed on BN ($\epsilon_0 = 2.63$, empty symbols): $\mathrm{MoS}_2$ (circles),  $\mathrm{WS}_2$ (squares), $\mathrm{MoSe}_2$ (triangles), 
$\mathrm{WSe}_2$ (diamonds). For each material there are two points due to the SOC.  
The material information is obtained from Ref.~\onlinecite{Xiao2012csa}.}
\label{fig:a_vs_S}
\end{figure}

The lowest energy levels, corresponding to $j = 1/2$, are the first to merge with the continuum. 
We determine the smallest values of $\alpha$ that satisfy Eq.~\eqref{eqn:alpha_sol} for a given $S_l$ at $j = 1/2$ 
and plot the result in Fig.~\ref{fig:a_vs_S}. We also show where four STMDC's are located in the
phase diagram depending on the dielectric screening. Suspended samples with $\epsilon_0 = 1$ are all located
in the supercritical regime. On the other hand, samples placed on BN with $\epsilon_0\approx 2.63$ are located 
in the subcritical regime. Therefore, in order to possibly observe the excitonic collapse, for $S_l \approx 1$ 
one needs to work substrates with $1 < \epsilon_0 \leq 1.25$. 

\section{Conclusion}
\label{sec:Conclusion}

In conclusion, we have analyzed a long wavelength model for STMDC's and obtained both the long-range and static polarization functions. 
We used the latter to determine the screening of Coulomb interactions in the system. This allowed us to solve for the energy levels of 
exciton quasiparticles. Most importantly, we obtained an equation which gives the critical coupling $\alpha$ in a system that leads to a 
so-called excitonic collapse. Our results indicate that for strong enough coupling this phenomenon should be omnipresent in this class of materials.

We acknowledge DOE grant DE-FG02-08ER46512, ONR grant MURI N00014-09-1-1063, and the NRF-CRP award 
"Novel 2D materials with tailored properties: beyond graphene" (R-144-000-295-281).

\end{document}